\documentclass[review]{elsarticle}

\usepackage{lineno,hyperref}
\modulolinenumbers[5]

\journal{International Journal of Multiphase Flow}

\usepackage{color}
\usepackage{multirow}
\usepackage{dcolumn}% Align table columns on decimal point
\usepackage{multicol}
\usepackage{graphicx}
\usepackage{caption}
\usepackage{amssymb}
\usepackage{epstopdf}
\usepackage{mathtools}
\usepackage{float}
\usepackage{bm}% bold math
\usepackage{comment}

\usepackage{subcaption}
\usepackage{graphicx}

\biboptions{authoryear}

\biboptions{sort&compress}
\usepackage{color}

\newcommand{\ks}{\textcolor{black}} % for commenting
\begin{document}

\begin{frontmatter}

%\title{Dynamics of Drop-on-Drop Impacts on Liquid Surfaces}
\title{Application of deep learning and inline holography to estimate the droplet size distribution}
\author{Someshwar Sanjay Ade$^a$, Deepa Gupta$^b$, Lakshmana Dora Chandrala$^c$ and Kirti Chandra Sahu$^{b}$\footnote{lchandrala@mae.iith.ac.in; ksahu@che.iith.ac.in} }
\address{$^a$Center for Interdisciplinary Program, Indian Institute of Technology Hyderabad, Kandi - 502 284, Sangareddy, Telangana, India \\ 
$^b$Department of Chemical Engineering, Indian Institute of Technology Hyderabad, Kandi - 502 284, Sangareddy, Telangana, India\\
$^c$Department of Mechanical and Aerospace Engineering, Indian Institute of Technology Hyderabad, Kandi - 502 284, Sangareddy, Telangana, India}

\begin{abstract}
We examine five machine learning-based architectures to estimate the droplet size distributions obtained using digital inline holography. The architectures, namely, U-Net, R2 U-Net, Attention U-Net, V-Net, and Residual U-Net are trained using synthetic holographic images. Our assessment focuses on evaluating the training, validation, and prediction performance of these architectures. We found that U-Net and R2 U-Net to be the most proficient, displaying consistent performance trends and achieving the highest Intersection Over Union (IOU) scores compared to the other three architectures. We employ additional training using experimental holographic images for the two top-performing architectures to validate their efficacy further. Subsequently, they are employed to segment an experimental dataset illustrating the bag breakup phenomenon, facilitating the extraction of size distribution. The extracted size distribution from U-Net and R2 U-Net segmentation is then compared with the analytical model proposed by \cite{jackiw2022prediction} by employing the gamma and log-normal distributions. Our findings indicate that the gamma distribution provides a more accurate prediction of the multi-modal size distribution than the log-normal distribution owing to its long exponential tail. The present study offers valuable insights into the effectiveness of machine learning architectures in estimating particle/droplet sizes, highlighting their practical application in real-world experimental scenarios.
\end{abstract}
\end{frontmatter}


\section{Introduction} \label{sec:intro}

The fragmentation of droplets and the resulting distribution of smaller droplet sizes are important in various applications, including combustion, spray coating and deposition, pharmaceuticals, drug delivery, and modelling disease transmission \citep{villermaux2007fragmentation, marmottant2004spray, soni2020deformation, jackiw2021aerodynamic, kirar2022experimental, xu2022droplet}. Beyond their significance in industrial and medical contexts, child droplets in the form of aerosols also play a significant role in understanding natural phenomena, particularly influencing atmospheric processes such as cloud formation and precipitation \citep{Villermaux2009single, villermaux2011distribution}.

The analysis of satellite droplet size distribution resulting from fragmentation is commonly performed using Laser Diffraction (LD), Phase Doppler Particle Analyser (PDPA), and in-line holography techniques \citep{swithenbank1976laser,dumouchel2009light, gao2013quantitative,katz2010applications,kumar2019automated}.  The LD technique is based on Mie's scattering theory, which describes how spherical particles scatter light. Laser diffraction instruments operate on the principle that the diffraction pattern produced by particles is inversely related to their size. As droplets pass through a laser beam, the diffracted light pattern is captured and analysed to determine the size distribution of the droplets. The PDPA technique operates on the principle of interferometry and laser Doppler velocimetry. It analyses droplets by measuring the phase difference and frequency shift of scattered light from particles. This provides simultaneous measurements of droplet size and velocity. Both LD and PDPA techniques are applicable in scenarios like continuous spray atomisation. However, these methods suffer from restricted spatial resolution. Additionally, PDPA is limited to spherical particles, while LD provides an average drop size measurement along a designated line of sight.

Recent advancements in digital in-line holography (DIH) have demonstrated its effectiveness as a reliable method for acquiring detailed three-dimensional information about particles \citep{shao2020machine,katz2010applications}. In contrast to point measurement approaches such as the LD and PDPA techniques, the DIH operates as a comprehensive, whole-field particle size measurement technique. The DIH technique utilises a coherent light source, such as a laser, and a single camera to capture the interference pattern. This pattern results from the interaction between the scattered light from the particles and the unscattered part of the illumination light source \citep{katz2010applications}. Traditionally, the hologram is numerically reconstructed, and subsequently, the size and location information of particles within the hologram is obtained through a segmentation process applied to the reconstructed optical field.

The primary challenge in the DIH technique lies in the segmentation of particles after reconstruction from the hologram. Numerous strategies have been proposed in the literature to tackle this challenge. For instance, \cite{tian2010quantitative} utilised a Gaussian mixture model for segmentation to determine the size distribution of bubbles in a well-mixed water tank. They also employed a minimum intensity matrix on the particle edge in the holograms to determine particle depth. Similarly, \cite{sentis2018bubbles} adopted a comparable approach to differentiate between bubbles and oil droplets in holograms.
On the other hand, several researchers  \citep{gao2014development, yingchun2014wavelet} utilised a minimum intensity metric for particle size measurement and a pixel intensity metric for measuring particle depth. This methodology has found application in spray measurement within a wind tunnel, as demonstrated by \cite{kumar2019automated}. Additionally, \cite{yingchun2014wavelet} implemented a wavelet filter for the reconstructed hologram and used the resulting filtered image as the focus metric, successfully applying this approach to coal particle size distribution. Moreover, \cite{talapatra2012application} applied particle shape segmentation criteria, determining particle depth based on pixel intensity gradient calculated from a Sobel filter. \cite{li2017size} employed a comparable approach for measuring the size distribution of droplets generated by breaking waves impinging on an oil slick. Furthermore, \cite{shao2019hybrid} introduced an approach that combines the intensity focus and wavelet-based focus metrics to estimate particle size distribution across a wide range of sizes. This method identifies small particles based on pixels showing a prominent peak in the longitudinal intensity profile, while larger particles are identified by minimum intensity projection. Several other studies, such as \citep{gao2013quantitative, guildenbecher2017characterization}, apply a match probability method to the reconstructed hologram to obtain the size and location of particles.

In the segmentation of reconstructed holograms with noisy data, machine learning utilizing deep neural networks (DNNs) has proven to be a powerful tool \citep{barbastathis2019use}. In the context of particle analysis, \cite{ilonen2018comparison} demonstrated that the implementation of CNN (Convolutional Neural Network) can provide higher accuracy in segmentation compared to methods such as watershed segmentation and intensity thresholding. While many holography studies employing machine learning for segmentation have focused on image modality transformations \citep{liu2019deep} and intensity and phase reconstruction of holograms \citep{rivenson2018phase,wang2018eholonet}, a subset of researchers \citep{hannel2018machine,ren2018learning,jaferzadeh2019no} has utilised a learning-based regression approach for segmenting reconstructed holograms containing a single object.

In machine learning, various neural network architectures are widely employed for image segmentation, with notable examples including U-Net, R2 U-Net, attention U-Net, V-Net, and residual U-Net. \ks{U-Net is a convolutional neural network architecture extensively utilised for semantic segmentation tasks \citep{ronneberger2015u}. It features a U-shaped structure with contracting and expansive paths to capture context and spatial information. U-Net can categorise each pixel in an image into a specific class or object. Moreover, the  U-Net architecture excels when there is limited training data, which is common in holographic image segmentation. R2 U-Net, proposed by \cite{alom2018recurrent} as an enhancement over U-Net, incorporates a recurrent residual module into the upsampling path. This augmentation facilitates the capture of broader context and refined information propagation across layers, which is essential for precise segmentation in holographic image analysis. Attention U-Net integrates self-attention mechanisms, allowing selective focus on relevant image regions and enhancing feature extraction. V-Net, tailored for medical image segmentation, employs volumetric convolutions to capture spatial dependencies \citep{milletari2016v} effectively. Residual U-Net combines the advantages of residual connections and U-Net, contributing to improved training convergence and feature learning \citep{guo2021channel}.} Recently, several studies \citep{shao2020machine,ade2022droplet} have applied the U-Net architecture in machine learning for holographic segmentation to determine size distribution. To the best of our knowledge, there has been limited exploration into different architectures for the segmentation of holographic images to extract particle size distribution.
 
In the present study, two crucial aspects are investigated. Initially, we determine the efficacy of advanced semantic segmentation architectures, including U-Net, R2 U-Net, Attention U-Net, V-Net, and Residual U-Net, in accurately determining the size of droplets/particles from reconstructed holograms. We evaluate these architectures in terms of their training, validation, and prediction performance. The optimised architecture is then tested against experimental data relating to bag breakup. Additionally, we extend our analysis by comparing these experimental findings with the analytical model developed by \cite{jackiw2022prediction}. The performance of the analytical model using Gamma-normal and Log-normal distributions is compared with experimental size distributions.

\section{Methodology}\label{sec:method}

The machine learning algorithm requires multiple training datasets, specifically involving reconstructed holograms and their corresponding binary masks. In \ks{the} following section, we discuss the generation of synthetic holograms, holographic reconstruction, and the processes involved in training the model. Furthermore, we evaluate the model using an experimental dataset related to droplets undergoing bag breakup. The description of the experimental setup, encompassing the shadowgraphy and holography systems, is provided below.

\subsection{Generation of synthetic hologram} \label{Generate_synthetic_images}

We adopt the methodology outlined by \cite{latychevskaia2015practical} for generating synthetic holograms, which is concisely summarised as follows. The process commences with calculating the Fourier transform of the object transmission function, representing the interaction of light with an object. Subsequently, the simulation of the Fourier transform of the Fresnel function is conducted to capture wavefront information during propagation. The outcomes of these two steps are multiplied, combining the spatial and spectral characteristics of the object. Following this multiplication, the holographic information is reconstructed by computing the inverse Fourier transform of the combined result. The final step involves obtaining the intensity distribution on the hologram by taking the square of the absolute value of the inverse Fourier transform.

\begin{figure}
\centering
\includegraphics[width=0.95\textwidth]{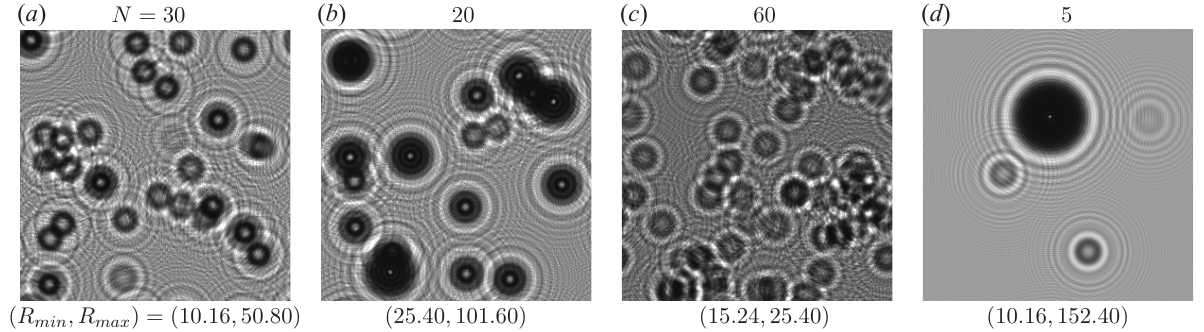}
\caption{A sample of synthetic holograms featuring spherical particles. The number of particles $(N)$ is indicated at the top of each panel. The minimum ($R_{\textrm{min}}$) and maximum ($R_{\textrm{max}}$) particle sizes in $\mu \textrm{m}$ are specified at the bottom of each panel. Within each panel, the particles follow a standard normal distribution in size and exhibit a random distribution in depth.}
\label{fig:3}
\end{figure}

The synthetic holograms are generated with a resolution of 5.08 $\mu \textrm{m}/\textrm{pixels}$, and each hologram has a size of $256\times 256$ pixels. Figure \ref{fig:3} shows sample images of synthetic holograms for spherical particles/droplets, generated using the described approach. The particles in the synthetic hologram adhere to a standard normal distribution in size and exhibit a random distribution in depth. The smallest and largest particle sizes ($R_{\textrm{min}}$ and $R_{\textrm{max}}$) in the synthetic hologram are 10.16 $\mu \textrm{m}$ and 152.40 $\mu \textrm{m}$, respectively. A total of 15 synthetic holograms are generated, each containing a minimum of 5 particles and a maximum of 60 particles.

\subsection{Holographic reconstruction and training dataset} \label{reconstruction}
\begin{figure}
\centering
\includegraphics[width=0.85\textwidth]{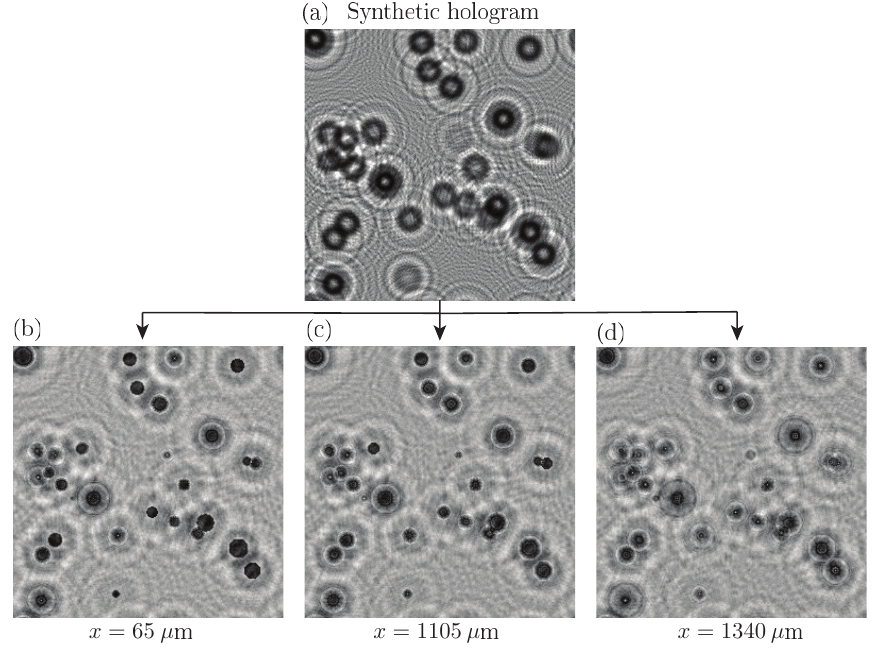}
\caption{Numerical reconstruction using the Rayleigh-Sommerfeld equation. (a) Depicts a synthetic hologram, and (b), (c), and (d) show the reconstructed images at different depths.}
  \label{fig:holo_recon}
\end{figure}

The synthetic holograms are numerically reconstructed using the Rayleigh-Sommerfeld equation, given by
\begin{equation}
I_{r}(x,y,z)=I_{h}(y,z)\otimes h(x,y,z),
\end{equation}
where $I_{r}(x,y,z)$ represents the 3D complex optical field obtained from reconstruction. The term $I_{h}(y,z)$ denotes the synthetic hologram, $\otimes$ signifies the convolution operation, and $h(x,y,z)$ is the diffraction kernel. The Rayleigh-Sommerfeld diffraction kernel in the frequency domain is expressed as \citep{katz2010applications}:
\begin{equation}
H(f_{x},f_{y},z)=\textrm{exp}\left (jkz\sqrt{1-\lambda^{2}f_{x}^{2}-\lambda^{2}f_{y}^{2}}\right),
\end{equation}
wherein $j = \sqrt{-1}$ and $k = 2\pi/\lambda$ represent the imaginary unit and the wavenumber, respectively; $\lambda$ is the wavelength of the incident beam. The spatial frequencies in the $x$ and $y$ directions are denoted by $f_{x}$ and $f_{y}$, respectively. In the frequency domain, using the convolution theorem, the complex optical field (3D image of objects), $I_{r}$ can be evaluated as
\begin{equation}
I_{r}(x,y,z)={\rm FFT}^{-1} \left \{{\rm FFT} \left [I_{h}(y,z)\right]\times H(f_{x},f_{y},z)\right \}.
\end{equation}
Here, the operator $\text{FFT}$ denotes the Fast Fourier Transform. The intensity information within a given plane is derived from the magnitude of the complex optical field in that specific plane. The reconstruction process is executed across a sequence of planes, each separated by a depth spacing of $1~\mu$m. The dimensions of the reconstructed volume measure $1.3$ mm in each direction. Figure \ref{fig:holo_recon} depicts the reconstruction results, showcasing both the synthetic holographic and reconstructed images at various depths ($x$).

The dataset is created for model training using images obtained through the reconstruction process and their corresponding binary masks of particles/droplets. In order to enhance the diversity of the dataset, various data augmentation techniques, such as translation, rotation, and shear deformation, are applied to the reconstructed images. \ks{These augmentations are crucial in capturing variations in droplet orientation, position shifts, and deformations observed in holographic images. By incorporating these techniques, we ensure the robustness of the model in accurately predicting droplet sizes.} A detailed illustration of the training process is provided in the subsequent section.

\subsection{Model training} \label{model_training}

We outline the training process for five distinct architectural configurations: U-Net, R2 U-Net, Attention U-Net, V-Net, and Residual U-Net. A set of 500 droplet images, generated through synthetic hologram reconstruction and subsequent data augmentation, is employed to train these configurations. \ks{The augmented dataset enhances the capability of the model to capture variations and patterns within the data, resulting in improved generalisation and performance on new or unseen data \citep{frid2018synthetic}.} The training process also utilises the corresponding ground truth masks for these 500 images. These masks serve as pixel-wise classifications that the network aims to predict as the optimal outcome. The training procedure was executed using Keras version 2.11.0 on an NVIDIA RTX A5500 GPU. Before inputting the images into the architectural configurations, normalisation is applied to scale the dataset within the range of 0 and 1. The hyperparameter values, such as the learning rate, epochs, and batch size used in our training algorithm, are set to 0.001, 800, and 10, respectively. The dataset is partitioned into training and validation sets in an 8:2 ratio during the training process. These sets are then used to feed the models, and their corresponding performance metrics are computed \citep{alpaydin2020introduction}. 

The training accuracy is defined as the ratio of the number of correct predictions in the training dataset to the total number in the training dataset. Similarly, the validation accuracy is defined as the ratio of the number of accurate predictions in the validation dataset to the total number in the validation dataset. The Intersection over Union (IOU) is defined as the area of overlap $({P \cap Q})$ divided by the area of union $({P \cup Q})$, where $P$ and $Q$ represent the predicted segmentation map and the actual ground truth, respectively.

The intersection over union (IOU) metric, also known as the Jaccard Index, is utilised to assess the predicted segmentation against the ground truth, producing values between 0 and 1. This metric measures the similarity between two sets of samples, where a score of 0 signifies no overlap, and a score of 1 indicates complete and identical overlap. To refine the model's precision, we conducted multiple iterations of the training process over 800 epochs, adjusting hyperparameters to evaluate their impact on performance. The model aims to minimise binary cross-entropy loss, employing the Stochastic Gradient Descent (SGD) optimisation algorithm to update weights after each batch. SGD, a modified version of Gradient Descent, is designed to enhance the performance of machine learning models by addressing computational challenges associated with handling large datasets. This comprehensive training routine enabled us to improve model accuracy by identifying optimal parameter values and achieving convergence.

The architectures incorporate max pooling layers to downsample feature maps, reducing their spatial dimensions progressively. This non-linear downsampling focuses on the most activated features, decreasing computational requirements for subsequent layers and simultaneously serving as a form of regularisation to mitigate overfitting. Following the bottleneck, transposed convolutions are employed to upsample feature maps back to the original input resolution. This un-pooling process reconstructs high-resolution activations by interpolating the spatially reduced outputs from max pooling, resulting in a symmetric contraction and expansion of the feature representation.
The leaky rectified linear unit (Leaky ReLU) is applied as an activation function for hidden layers, enhancing computational speed during the training process, preventing the occurrence of the "dying ReLU" problem, and allowing for a small non-zero gradient for negative inputs compared to the use of the traditional ReLU activation function ($f(x)$). The model utilises the sigmoid activation function in the output layer to generate probability predictions for classifying pixels as background or droplets. This commonly used classifier maps any real value to a probability between 0 and 1. Applying the sigmoid activation function enables the network to produce likelihood predictions that a given pixel belongs to the droplet class based on the learned features. The activation function, $f(x)$ is given by
\begin{equation}
\centering
 f(x) = \begin{Bmatrix}
 0.01x & {\rm for}~x< 0 \\ 
 x& {\rm for}~x\geq 0
\end{Bmatrix},
\end{equation}
where $x$ denotes the weighted sum of inputs into a node or artificial neuron.

In the present study, the segmentation task involves classifying pixels into droplets or non-droplets. Given our focus on binary semantic segmentation of droplets, we framed the problem as binary pixel-wise classification. Consequently, we chose binary cross-entropy as the primary loss function for model optimisation. Mathematically, the binary cross-entropy loss for each pixel can be expressed as \citep{alpaydin2020introduction}:
\begin{equation} 
\centering
H=- \frac{1}{n}\sum_{i=1}^{n}(y_{i}\textrm{log}(\hat{y}_{i})+(1-y_{i})\textrm{log}(1-\hat{y}_{i})),
\end{equation}
where $H$ is the binary cross entropy loss. $n$ is the number of images. $y_i$ is the ground truth, and $\hat{y}_{i}$ is the actual output of the model network.

\subsection{Experimental setup} \label{sec:exp}

\begin{figure}
\centering
\includegraphics[width=0.9\textwidth]{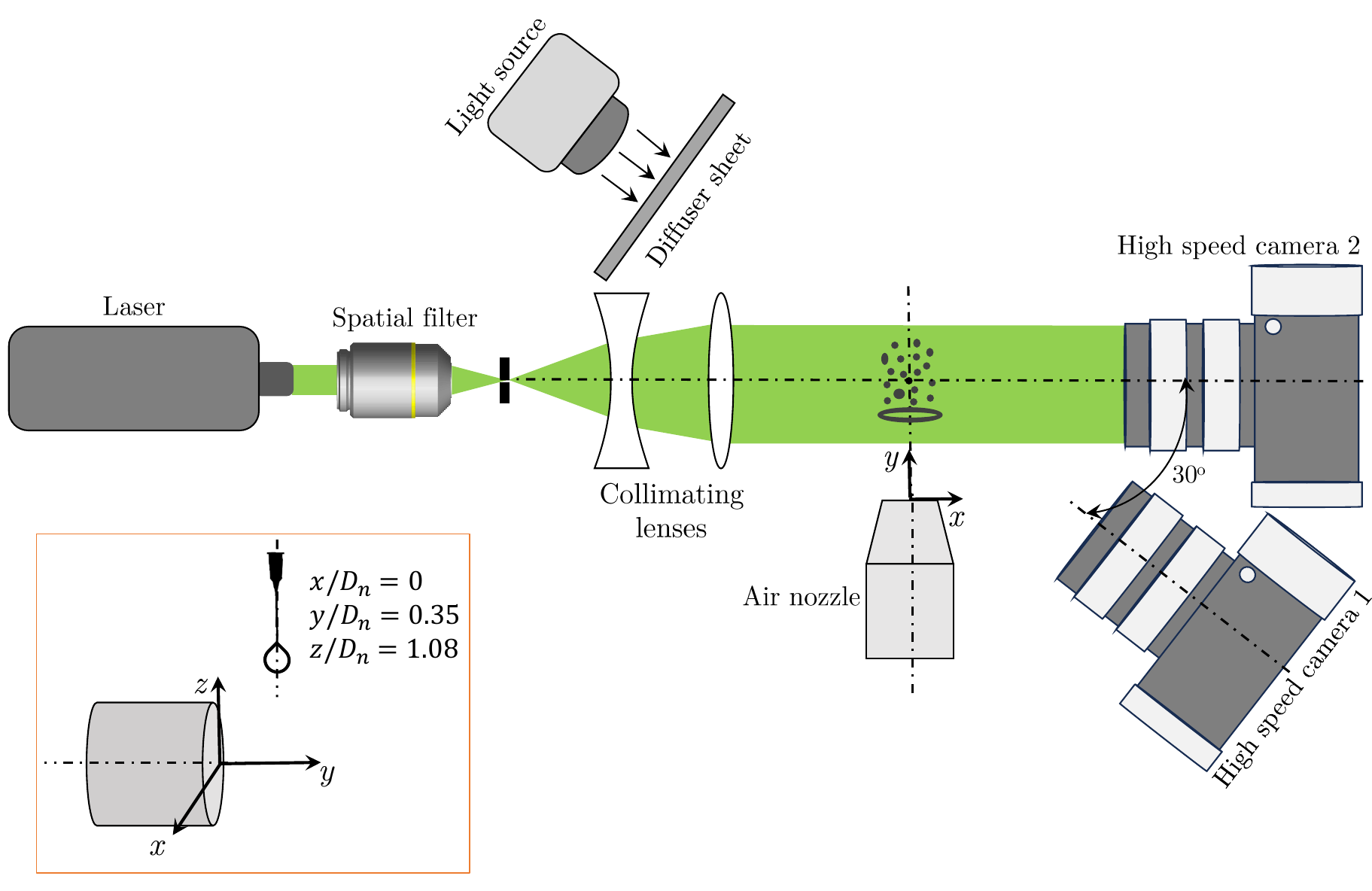}
\caption{Schematic of the experimental setup (top-view) involving shadowgraphy and digital in-line holography to acquire the size distribution of child droplets resulting from the breakup of a parent droplet. The inset provides the positioning of the dispensing needle with respect to the centre of the air nozzle, where $D_n$ represents the diameter of the air nozzle, set at 18 mm.}
\label{exp_setup}
\end{figure}

We employ a dual technique setup involving shadowgraphy to observe the morphology of a droplet undergoing fragmentation and digital in-line holography to analyse the size distribution of child droplets resulting from this breakup. The experimental arrangement includes (i) an 18 mm diameter air nozzle, (ii) a droplet dispensing needle, (iii) a continuous wave laser, (iv) a spatial filter arrangement, (v) collimating optics with concave and convex lenses, (vi) two high-speed cameras, and (vii) diffused backlit illumination. Figure \ref{exp_setup} illustrates a schematic diagram of the complete experimental setup.

A Cartesian coordinate system $(x, y, z)$ is used to describe the dynamics, with its origin situated at the centre of the air nozzle as shown in Figure \ref{exp_setup}. The dispensing needle is positioned at $(x/D_{n}, y/D_{n}, z/D_{n}) = (0, 0.35, 1.08)$. The high-speed camera 1 is dedicated to shadowgraphy and is located at $x=180$ mm, making an angle of $-30^\circ$ with the $x$ axis. A high-power light-emitting diode, combined with a uniform diffuser sheet, is used for background illumination. The images captured by high-speed camera 1 at 1800 frames per second (fps), with an exposure duration of 1 $\mu$s and a spatial resolution of 31.88 $\mu$m/pixel, have a resolution of $2048 \times 1600$ pixels.

A continuous wave laser, with an output power of 100 mW and a wavelength of 532 nm, along with a spatial filter, collimating lenses, and high-speed camera 2 positioned at $x=180$ mm, as illustrated in Figure \ref{exp_setup}, are employed for digital in-line holography. The spatial filter comprises an infinity-corrected plan achromatic objective (20X magnification) and a 15 $\mu$m pin-hole to generate a clean beam. This beam is expanded using a plano-concave lens and collimated using a plano-convex lens, illuminating the droplet field of view. The resulting interference patterns from the child droplets are captured by high-speed camera 2 at a resolution of $2048 \times 1600$ pixels. The camera records at 1800 fps with an exposure duration of 1 $\mu$s and a spatial resolution of 15.56 $\mu$m/pixel.

\section{Results and discussion}\label{sec:dis}

The presentation of our results begins with an analysis of the performance of five different architectures trained with 500 synthetic images. The evaluation encompasses testing performance during the training, validation, and prediction. After this initial assessment, the two best-performing architectures undergo additional training using experimental images of droplet breakup. Ultimately, the trained architectures are deployed to segment the experimental data, enabling the determination of the size distribution of droplets.

\subsection{Performance of the model trained using synthetic images}

We utilise R2 U-Net, U-Net, Attention U-Net, V-Net, and Residual U-Net architectures, and the resulting performance parameters are presented in Table \ref{table:T2}. A detailed examination of these parameters reveals that R2 U-Net and U-Net perform superiorly among the five architectures. It can be observed that the performance metrics for R2 U-Net and U-Net closely align, except for the validation IOU.

\begin{table}
\centering
\begin{tabular}{|c|c|c|c|c|c|}
\hline
Architecture      & R2 U-Net & U-Net & Attention  & V-Net & Residual  \\ 
   &  & & U-Net & V-Net & U-Net \\ \hline
Loss  & 0.0096  & 0.0098  & 0.0102   & 0.0129 & 0.0152 \\ \hline
Accuracy  & 0.9900 & 0.9900 & 0.9898  & 0.9889 & 0.9885 \\ \hline 
IOU  & 0.8543  & 0.8524 & 0.8469  & 0.8070 & 0.7425 \\ \hline
Validation loss  & 0.0101  & 0.0169 & 0.0273 & 0.0204  & 0.0230 \\ \hline
Validation accuracy & 0.9959 & 0.9936 & 0.9911 & 0.9921 & 0.9899 \\ \hline
Validation IOU & 0.8636 & 0.8241  & 0.7632 & 0.7668 & 0.7389 \\ \hline
\end{tabular}
\caption{Comparison of performance parameters obtained using input data set of 500 synthetic images for different architectures.}
\label{table:T2}
\end{table}

\begin{figure}
\centering
\includegraphics[width=1\textwidth]{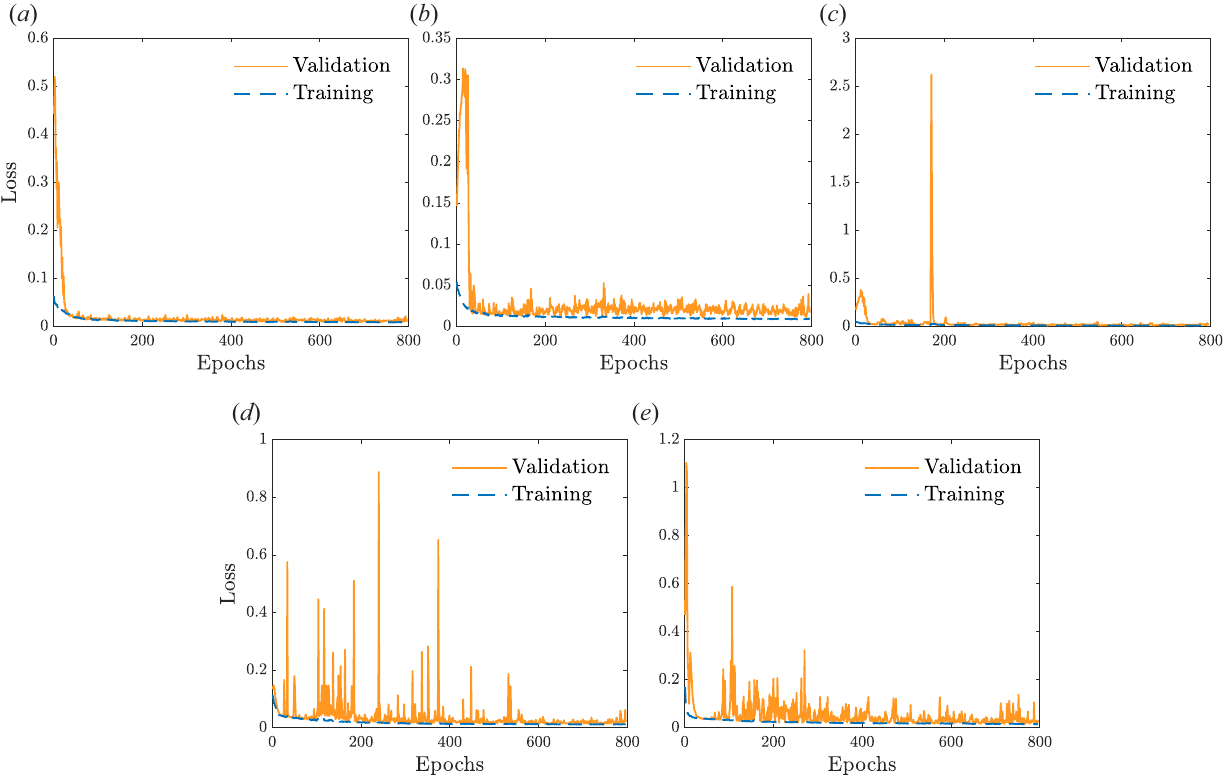}
\caption{The variation of the training loss and validation loss with the number of epochs obtained using a data set of 500 synthetic images for different architectures. (a) R2 U-Net, (b) U-Net, (c) Attention U-Net, (d) V-Net and (e) Residual U-Net.}
\label{fig:1}
\end{figure}

\begin{figure}
\centering
\includegraphics[width=1\textwidth]{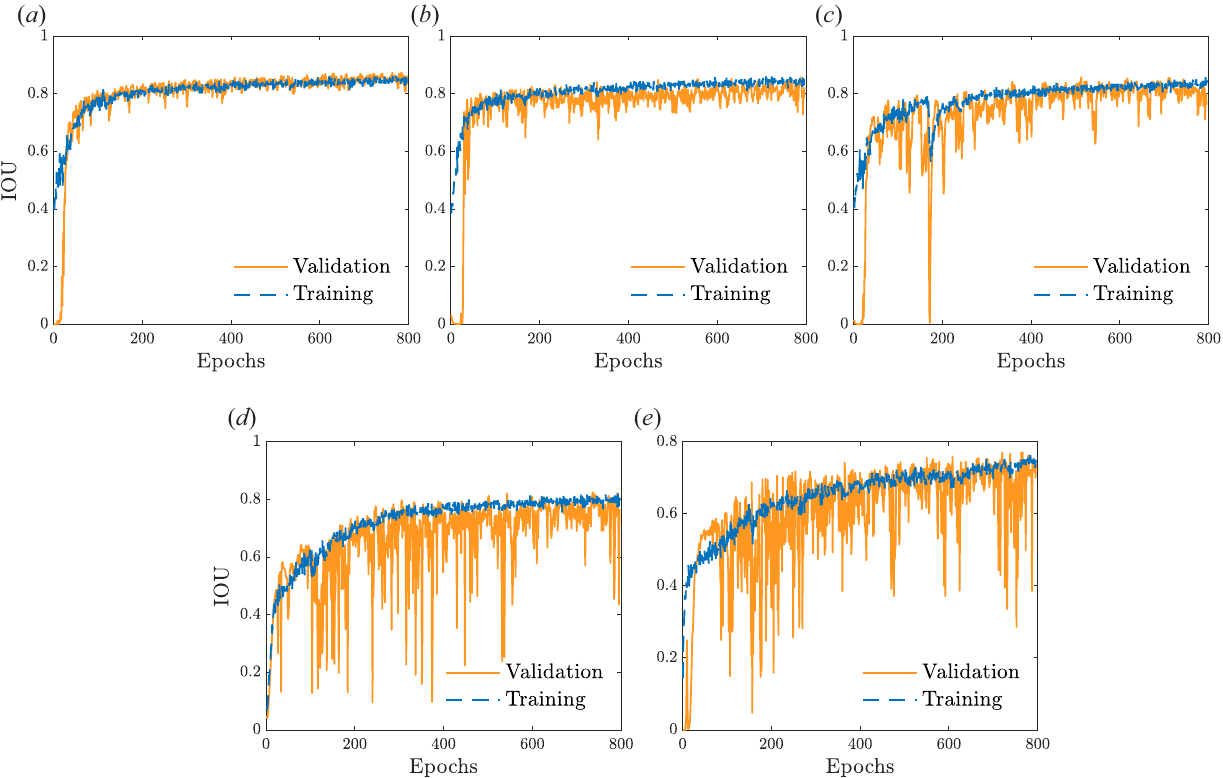}
\caption{The evolution of intersection over union (IOU) across epochs for various architecture-trained models using a synthetic dataset comprising 500 images. The architectures include (a) R2 U-Net, (b) U-Net, (c) Attention U-Net, (d) V-Net, and (e) Residual U-Net.}
\label{fig:2}
\end{figure}

Figure \ref{fig:1} depicts the variation in training and validation loss across different architectures with respect to the number of epochs. It is evident that both R2 U-Net and U-Net consistently demonstrate a stable performance trajectory in terms of validation loss during training. Conversely, the Attention U-Net, V-Net, and Residual U-Net architectures exhibit a more pronounced noise component and large sharp jumps in validation loss, indicating undesirable features in their training history. Figure \ref{fig:2} shows the IOU variation with epochs. It can be seen that R2 U-Net and U-Net architectures outperform other architectures, showcasing the highest IOU scores. These architectures also display a more stable rise in validation IOU throughout training compared to others. Conversely, Attention U-Net, V-Net, and Residual U-Net experience intensified fluctuations with sharp drops in validation IOU, as depicted in Figures \ref{fig:2}(c-e), respectively.

Next, we evaluate the predictive performance of the two most effective architectures in identifying droplets within an image. During this assessment, the architectures make predictions for droplet regions, which are then compared to the ground truth (i.e., mask). The predicted droplet regions are assigned a value of 1, while the remaining areas are assigned a value of 0, forming an image based on these pixel values. Figure \ref{fig:drop_comp} compares droplet segmentation results obtained from the U-Net and R2 U-Net architectures against the ground truth data. As illustrated in the figure, both architectures efficiently detect droplet regions within the background. Figure \ref{fig:drop_comp}(d) and (e) clearly show that the droplet size distribution, as determined by the number probability density ($P_n$), closely aligns with the ground truth data for both U-Net and R2 U-Net.

\begin{figure}
\centering
\includegraphics[width=0.9\textwidth]{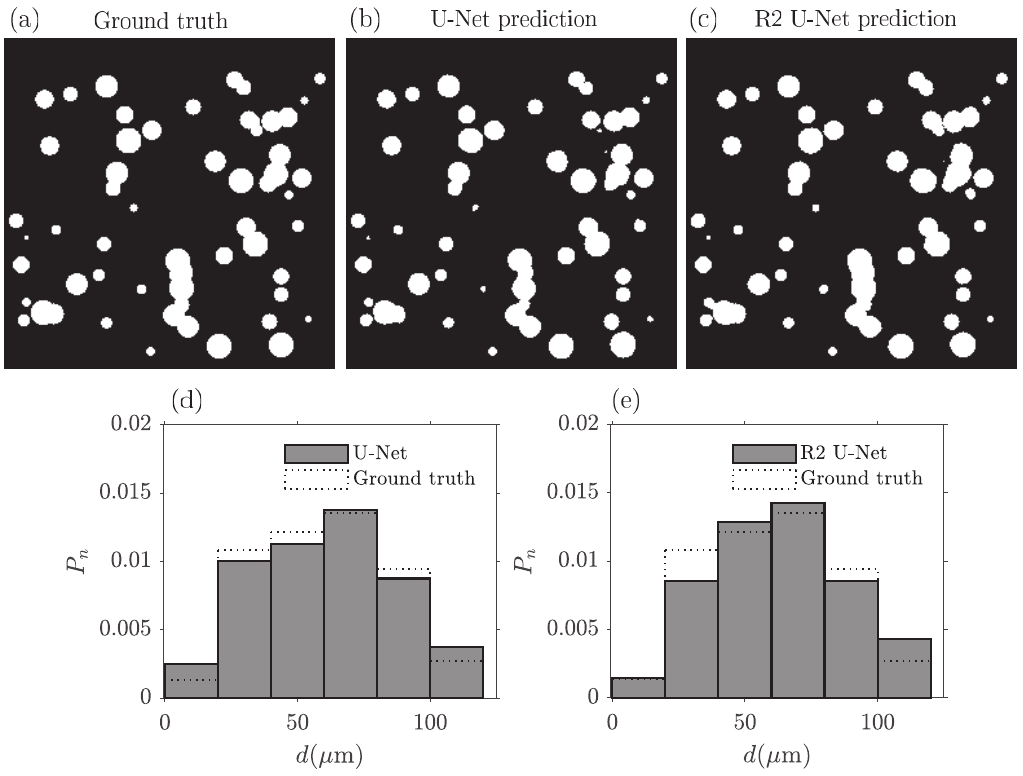}
\caption{Comparison between the ground truth and the measured distribution from the U-Net and R2 U-Net architectural configurations for a dataset consisting of 60 droplets randomly distributed in depth. (a) The ground truth image. (b) and (c) The predicted results from the U-Net and R2 U-Net architectures, respectively. Panels (d) and (e) illustrate the size distribution comparison between the U-Net and R2 U-Net architectures with the ground truth data. Here, $d$ and $P_n$ represent the diameter of the droplets and the number probability density, respectively.}
\label{fig:drop_comp}
\end{figure}

\subsection{Model trained with experimental images}

\begin{figure}
\centering
\includegraphics[width=0.8\textwidth]{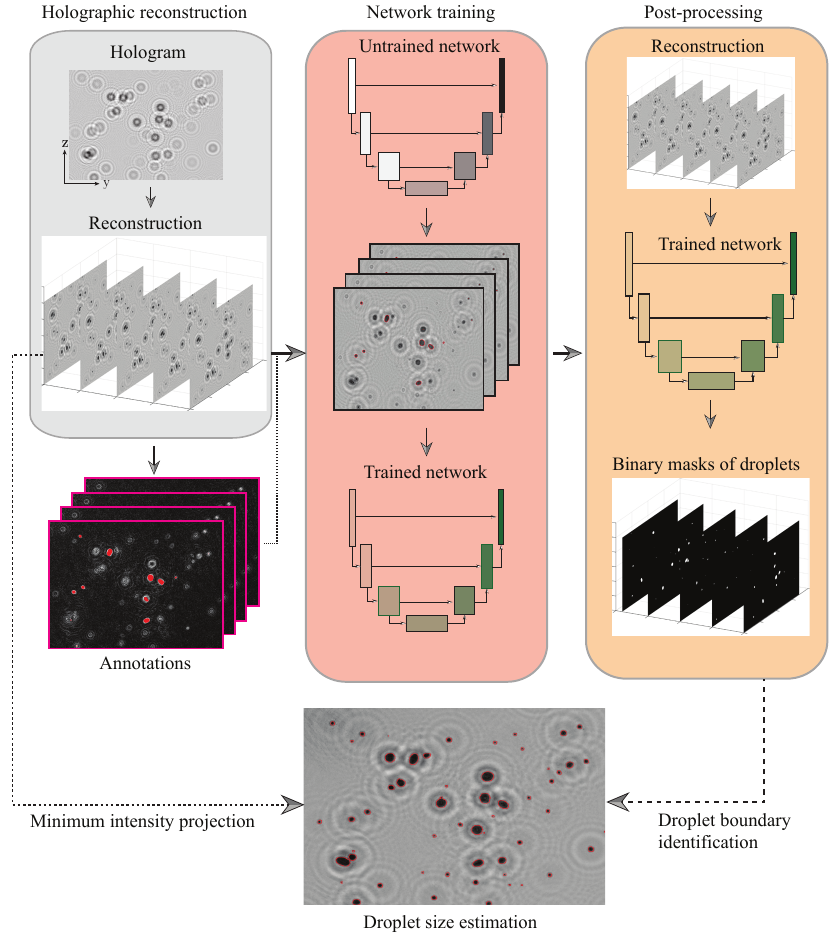}
\caption{Sequential steps involved in processing the hologram obtained from digital inline holography experiments.}
\label{fig:5}
\end{figure}

After successfully training the model architectures using 500 synthetic images, each sized at $256\times256$ pixels, we train the models using experimental images. Fifty experimental images, also sized at $256\times256$ pixels each, are employed for this phase. Based on the insights gained from the performance of R2 U-Net and U-Net on synthetic images, we exclusively utilise these two architectures to train and assess their performance on the experimental image dataset. \ks{The experimental images undergo two preprocessing steps before being utilised for training. The background image is initially subtracted from the recorded holograms in the first step. Subsequently, in the second step, intensity normalisation is performed to eliminate noise and rectify uneven illumination present in the images. This normalisation process is expressed as $I_{N}=(I-I_{min})/(I_{max}-I_{min})$, where $I_{N}$ denotes the normalised image, and $I_{min}$ and $I_{max}$ represent the minimum and maximum intensities in the image, respectively. Additionally, 30 holograms without particles are recorded before each experiment commences, which is then averaged to obtain the mean intensity image of the background.} Figure \ref{fig:5} illustrates the sequential stages of pre-training, training, and post-training processing images obtained from digital inline holography experiments. The process commences with hologram reconstruction using the Rayleigh Sommerfeld equation, following the methodology outlined for synthetic holograms in Section \ref{reconstruction}. The subsequent step involves training the model for two architectural configurations, \ks{namely}, U-Net (Figure \ref{fig:5}a) and R2 U-Net (Figure \ref{fig:5}b), to segment droplet images. Before entering the training phase, the reconstructed images undergo data augmentation. \ks{The model iterates through multiple epochs using the training dataset in the training process. Within each epoch, batches of augmented data are presented to the model. It computes the loss by comparing its predicted outputs with the actual ground truth. Then it adjusts its weights using backpropagation and gradient descent to minimise this loss. Across successive epochs, the model progressively acquires the ability to identify the underlying patterns within the data, thus improving its performance over time. Additionally, validation data is employed to evaluate the performance of the model on unseen data and mitigate overfitting. After each epoch, the model assesses its performance on the validation dataset, computing metrics like loss and accuracy.}

In the current investigation, a set of 50 experimentally obtained images from the reconstructed volume, each manually annotated as ground truth, is employed for network training. The ground truth annotations are carried out through local thresholding around each satellite droplet. The final step involves post-processing of the 3D reconstructed volume to determine droplet boundaries. The trained network is applied to each plane within the reconstructed 3D volume, as illustrated in Figure \ref{fig:5}. The output of the network directly provides binary masks corresponding to droplet boundaries in each plane. The ultimate processing steps include maximum intensity projection of the binary mask, removal of droplets smaller than 3 pixels, and the estimation of diameters of the remaining droplets.

Table \ref{table:T3} presents a comparative analysis of performance metrics for the experimental image-based training. Notably, the training loss, accuracy, and IOU metrics demonstrate closely aligned values for both architectures. This coherence extends to trends observed in validation loss, validation accuracy, and validation IOU as well. In this context, the difference between R2 U-Net and U-Net validation IOU has decreased. Nevertheless, R2 U-Net exhibits a slight performance advantage over the U-Net architecture, underscoring its effectiveness in handling experimental image data.

\begin{table}
\centering
\begin{tabular}{|c|c|c|}
\hline
Parameters       & R2 U-Net & U-Net \\ \hline
Loss                & 0.0093            & 0.0091          \\ \hline
Accuracy            & 0.9883            & 0.9883          \\ \hline
IOU                 & 0.9093            & 0.9109          \\ \hline
Validation loss     & 0.0651            & 0.0745          \\ \hline
Validation accuracy & 0.9876            & 0.9867          \\ \hline
Validation IOU      & 0.8400            & 0.8292          \\ \hline
\end{tabular}
\caption{Comparison of performance parameters obtained using input data set of 50 experimental images for R2 U-Net and U-Net architectures.}
\label{table:T3}
\end{table}

\subsection{Prediction of size distribution of child droplets}

In this section, we demonstrate the size distribution of child droplets resulting from the fragmentation of a parent droplet under an airstream using U-Net and R2 U-Net architectures. Furthermore, we compare these experimental findings with the analytical model developed by \cite{jackiw2022prediction}. Additionally, we apply this analytical model for droplet size prediction, considering both Gamma and log-normal distributions. These distributions are commonly used to characterise droplet size distribution in drop fragmentation and spray.
\begin{figure}
\centering
\includegraphics[width=0.9\textwidth]{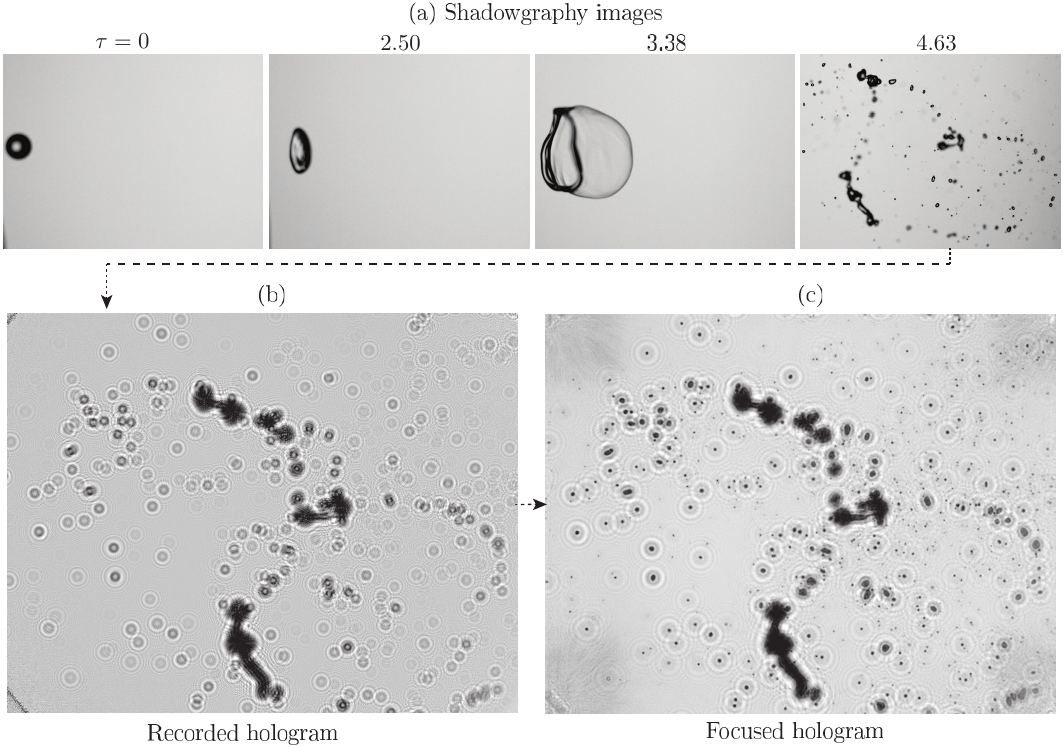}
\caption{Temporal evolution of the morphology of the droplet in air airstream and undergoing fragmentation. Panel (a) shows shadowgraphy images, while panel (b) presents the recorded hologram at $\tau= 4.63$. In panel (c), the depicts the corresponding focused hologram obtained after reconstruction, followed by minimum intensity projection from the recorded hologram in panel (b).}
\label{recorded_hologram_figv2}
\end{figure}

Figure \ref{recorded_hologram_figv2}(a) shows the temporal evolution of the morphology of the parent water droplet ($d_{0}=3.2$ mm) in an airstream and its subsequent fragmentation. For the flow rate of the airstream considered in the experiment, the Weber number is given by $We = \rho_{a} U^{2} d_{0}/\sigma = 17.5$, where $\rho_{a}$, $U$ and $\sigma$ represent the air density, average velocity of the airstream, and liquid surface tension, respectively. In Figure \ref{recorded_hologram_figv2}a, the dimensionless time is denoted as $\tau = Ut\sqrt{\rho_{a}/\rho}/d_{0}$ and is indicated at the top of each panel. Here, $t$ and $\rho$ refer to the physical time and density of water, respectively, and $\tau = 0$ signifies the moment when the freely falling parent droplet enters the potential core of the airstream. 
The fragmentation process of the droplet for various Weber numbers has been thoroughly investigated in \cite{ade2023size}.

In Figure \ref{recorded_hologram_figv2}a, it is evident that the initially spherical droplet (at $\tau=0$) undergoes deformation into a disk shape at $\tau=2.5$, driven by the influence of drag force over surface tension force. For this Weber number, with the progression of time ($\tau=3.38$), aerodynamic forces come into play as the parent droplet enters the potential core region of the airstream, causing the central region of the disk to elongate and form a thin liquid sheet (bag), accompanied by the development of a thick toroidal rim on the periphery. In the subsequent phase, the liquid sheet undergoes breakup due to Rayleigh-Taylor instability, and the rim fragments due to capillary instability. Finally, at $\tau=4.63$, the entire droplet undergoes fragmentation, marking the completion of the breakup process. Figure \ref{recorded_hologram_figv2}b illustrates the recorded hologram corresponding to the shadowgraphy image at $\tau=4.63$. Figure \ref{recorded_hologram_figv2}c presents the focused hologram achieved through reconstruction from the recorded hologram followed by minimum intensity projection. The reconstructed images from this hologram are subsequently segmented using the U-Net and R2 U-Net architectures to determine the size distribution.

\subsubsection{Prediction of droplet size using gamma distribution}\label{sec:gamma}

The volume probability density ($P_v$) is defined as the fraction of the total volume occupied by all child droplets of a particular diameter relative to the overall volume occupied by droplets of different diameters. Using the gamma distribution function, this is expressed as
\begin{equation} \label{j:eq1}
P_{v}=\frac{\zeta ^{3}P_{n}}{\int_{0}^{\infty}\zeta ^{3}P_{n}d\zeta}={\frac{\zeta ^{3}P_{n}}{\beta^{3}\Gamma (\alpha+3)/\Gamma (\alpha)}}.
\end{equation}
Here, $P_n$ is the number probability density function and can be evaluated as $P_n= {\zeta^{\alpha -1}e^{-\zeta /\beta} / \beta^{\alpha}\Gamma (\alpha)}$, where $\zeta \left (=d/d_0 \right)$ and $\Gamma (\alpha)$ represents the gamma function, wherein $\alpha=(\bar{\zeta}/\sigma_s)^{2}$ and $\beta=\sigma_s^{2}/\bar{\zeta}$ are the shape and rate parameters, respectively. The mean and standard deviation of the distribution are denoted as $\bar{\zeta}$ and $\sigma_s$ and are estimated based on the characteristic breakup sizes corresponding to each mode.

When the parent droplet undergoes a bag breakup, the associated fragmentation processes involve the node, rim, and bag modes. For an accurate representation of the size distribution, it is crucial to consider the contributions of each mode (node, rim, and bag). Typically, this is achieved by performing a weighted summation, where the contribution of each mode is multiplied by a weight factor that reflects its importance in the breakup process. Therefore, the total volume probability density ($P_{v,Total}$) can be calculated as:
\begin{equation} 
P_{v,Total}=w_{N}P_{v,N}+w_{R}P_{v,R}+w_{B}P_{v,B},
\end{equation}
\label{j:eq3}
where $w_{N}=V_{N}/V_{0}$, $w_{R}=V_{R}/V_{0}$ and $w_{B}=V_{B}/V_{0}$ represent the contributions of volume weights from the node, rim and bag, respectively. Here, $V_{N}$, $V_{R}$, $V_{B}$ and $V_{0}$ are the node, rim, bag and the initial droplet volumes, respectively. The volume probability density for the node, rim and bag breakup modes are denoted as $P_{v,N}$, $P_{v,R}$ and $P_{v,B}$, respectively. More details about the characteristic breakup sizes and the weight contributions for each mode are given in Appendix \ref{sec:App}. In the following, we present the size distribution of child droplets obtained from three repeated measurements for each set of parameters.

\begin{figure}
\centering
\includegraphics[width=0.95\textwidth]{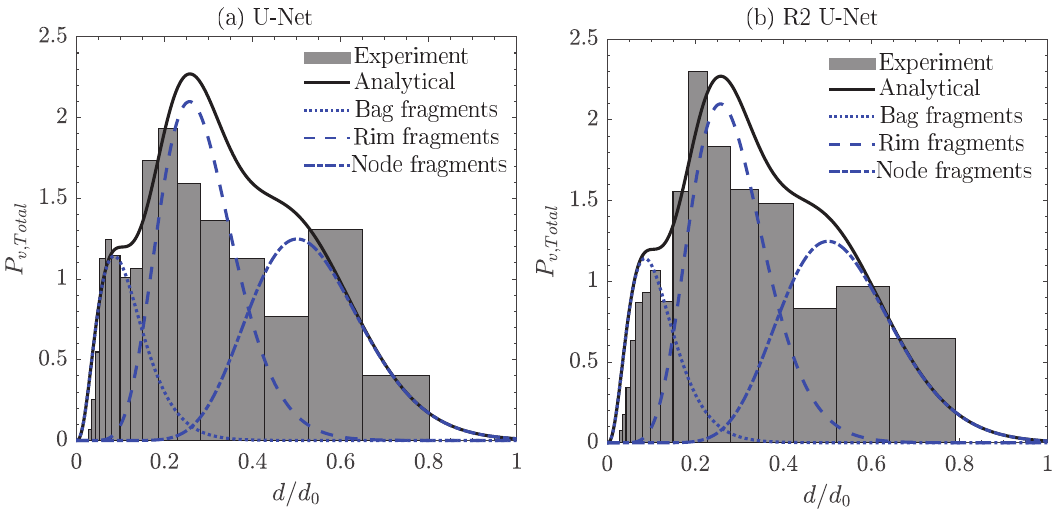}
\caption{Comparison of the analytical predictions of the droplet size distribution obtained using the gamma distribution function with experimental data segmented via (a) U-Net architecture and (b) R2 U-Net architecture.}
\label{gamma_dist}
\end{figure}

Figure \ref{gamma_dist}(a) shows the experimental size distribution of child droplets segmented using U-Net architecture for the bag breakup case at a typical instant $\tau=4.63$ and $We = 17.5$. The analytical prediction of the size distribution for individual modes and the overall size distribution are also depicted. In the bag breakup phenomenon, three distinct modes contribute to the overall size distribution: the first involves the bag rupturing due to the Rayleigh-Taylor (R-T) instability, the second comprises the rim fragmenting due to capillary instability, and the third entails the nodes breaking up due to R-T instability. In Figure \ref{gamma_dist}(a), the initial peak at $d/d_0 \approx 0.07$ is attributed to the bag rupture, while the subsequent peak at $d/d_0 \approx 0.20$ corresponds to the rim fragmentation. The third peak at $d/d_0 \approx 0.60$ signifies the node breakup. Additionally, Figure \ref{gamma_dist}(a) reveals that the analytically estimated contributions for bag, rim, and node breakups reasonably agree with the experimental size distribution related to these modes. Moreover, the theoretically predicted characteristic sizes align reasonably well with the corresponding experimental results. Notably, the volume fractions of bag, rim, and nodes are $15\%$, $45\%$, and $40\%$, respectively.

Figure \ref{gamma_dist}(b) depicts the size distribution of child droplets obtained through the segmentation via R2 U-Net architecture. Here, the first peak at $d/d_0 \approx 0.10$ is attributed to the bag rupture, while the second peak at $d/d_0 \approx 0.20$ corresponds to the rim fragmentation. The third peak at $d/d_0 \approx 0.57$ signifies the node breakup. Additionally, Figure \ref{gamma_dist}(b) also reveals that the analytically estimated contributions for bag, rim, and node breakups agrees well with experimental size distribution related to these modes.
 
\ks{In summary, the segmentation of experimental data utilising U-Net and R2 U-Net architectures for bag breakup yields a multimodal size distribution characterised by three distinct peaks (see Figure \ref{gamma_dist}(a) and \ref{gamma_dist}(b) of the manuscript). For U-Net (Figure \ref{gamma_dist}(a)), the first, second, and third peaks corresponding to bag, rim, and node distribution exhibit deviations from the analytical distribution, with percentage discrepancies of 9.6\%, 8.1\%, and 6\%, respectively. Conversely, for R2 U-Net (Figure \ref{gamma_dist}(b)), the first, second, and third peaks linked with bag, rim, and node distribution display deviations from the analytical distribution, with percentage discrepancies of 7\%, 9.5\%, and 29\%, respectively. These discrepancies underscore notable differences in the experimental size distributions following segmentation using U-Net and R2 U-Net architectures, respectively.}

\subsubsection{Prediction of droplet size using log-normal distribution}

\begin{figure}
\centering
\includegraphics[width=0.95\textwidth]{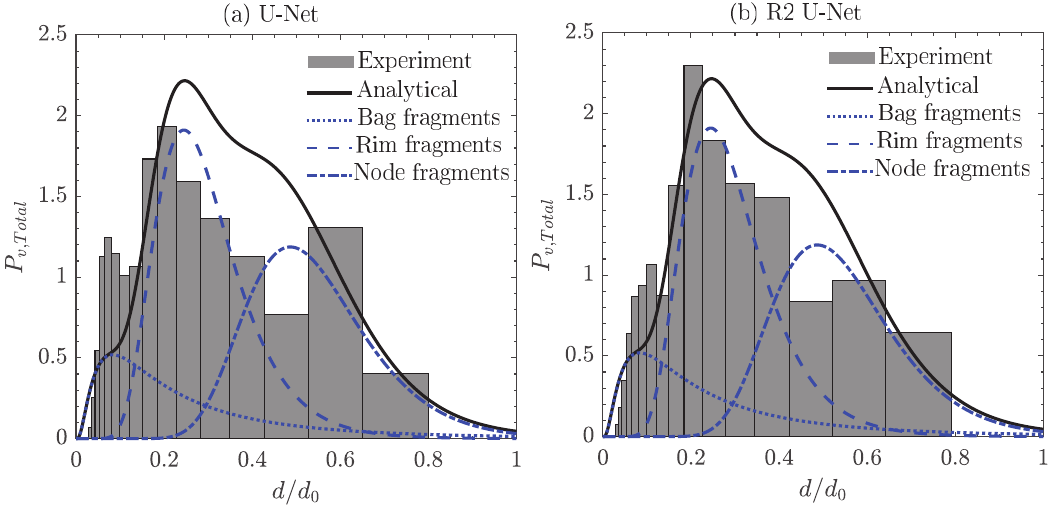}
\caption{Comparison of the analytical predictions of the droplet size distribution obtained using the log-normal distribution function with experimental data segmented via (a) U-Net architecture and (b) R2 U-Net architecture.}
\label{lognormal_dist}
\end{figure}

The volume probability density ($P_v$) for log-normal distribution is given by
\begin{equation} \label{j:aequ1}
P_{v}=\frac{\zeta ^{3}P_{n}}{\int_{0}^{\infty}\zeta ^{3}P_{n}d\zeta}={\frac{\zeta ^{3}P_{n}}{3\mu+4.5\sigma_{l}^2}},
\end{equation}
where, $\zeta \left (=d/d_0 \right)$ and $P_n$ is the number probability density function for log-normal distribution and it can be determined using following equation as 
\begin{equation} \label{j:aequ2}
P_{n}=\frac{1}{\zeta \sigma _{l}\sqrt{2\pi }}\textrm{exp}\left \{ \frac{-(\textrm{log}(\zeta)-\mu )^{2}}{2\sigma _{l}^{2}} \right \}.
\end{equation}
Here, $\mu $ and $\sigma_l$ are the logarithmic mean and logarithmic standard deviation of the distribution, which are given by
\begin{equation} \label{j:aequ3}
\mu =\textrm{log}\left ( \frac{\bar{\zeta }^{2}}{\sqrt{\sigma _{s}^{2}+\bar{\zeta }^{2}}} \right ) ~~ \textrm{and}~~
\sigma _{l}=\sqrt{\textrm{log}\left ( \frac{\sigma _{s}^{2}}{\bar{\zeta }^{2}}+1 \right )}.
\end{equation}
The process of estimating characteristic sizes and the overall size distribution for the log-normal probability density function is similar to that of the gamma distribution function, as discussed in Section \ref{sec:gamma}. Figures \ref{lognormal_dist}(a) and \ref{lognormal_dist}(b) depict the experimental size distribution of child droplets segmented using the U-Net and R2 U-Net architectures, respectively, at $\tau=4.63$ and $We=17.5$. Additionally, Figures \ref{lognormal_dist}(a) and \ref{lognormal_dist}(b) present the analytical predictions for individual modes and the overall size distribution obtained using the log-normal distribution function. It can be seen that the analytical distribution for the bag fragmentation mode under-predicts the corresponding experimental distributions obtained using U-Net and R2 U-Net architectures. This discrepancy is attributed to the long exponential tail inherent in the log-normal distribution. \ks{For U-Net (Figure \ref{lognormal_dist}(a)), the first, second, and third peaks corresponding to bag, rim, and node distribution exhibit deviations from the analytical distribution, with percentage discrepancies of 141.3\%, 1\%, and 10\%, respectively. Conversely, for R2 U-Net (Figure \ref{lognormal_dist} (b)), the first, second, and third peaks linked with bag, rim, and node distribution display deviations from the analytical distribution, with percentage discrepancies of 100.1\%, 20.4\%, and 22.6\%, respectively.} \ks{Furthermore, Figure \ref{fig:drop_comp} illustrates the comparison between the size distributions obtained using the conventional method (Hough transform) and the ground truth data. It is evident that the traditional method (Hough transform) exhibits inaccuracies in accurately segmenting the particles/droplets.}

\section{Conclusions} \label{sec:conc}

We systematically examine and compare the performance of five machine learning-based architectures, \ks{namely,} U-Net, R2 U-Net, Attention U-Net, V-Net, and Residual U-Net, to estimate the particle/droplet sizes obtained using digital inline holography. To ensure the robust training of these architectures, we create a synthetic dataset that includes numerous holograms with varying particle diameters and number densities. The evaluation of these architectures focuses on training and validation accuracy as well as IOU (Intersection over Union). The results emphasise U-Net and R2-U-Net as the most effective architectures, demonstrating superior accuracy and IOU scores compared to the other three architectures. We conduct digital inline holographic experiments involving a drop undergoing the bag breakup phenomenon. The performance of U-Net and R2-U-Net architectures is tested against the reconstructed holographic data from these experiments. Both U-Net and R2-U-Net successfully extract size information from the experimental data. We observe that the child droplets resulting from bag breakup exhibit a tri-modal distribution, indicating the occurrence of three distinct physical processes involving bag, rim and node fragmentation.  

Subsequently, we compare the experimental size distributions with the predictions from an analytical model utilising gamma and log-normal distribution functions. These distribution functions are commonly employed to comprehend the size distribution of child droplets in drop breakup and spray research. Our findings reveal that the gamma distribution offers a more accurate prediction of the intricate multi-modal size distribution resulting from bag fragmentation for the set of parameters considered. It is observed that while the gamma distribution function accurately captures the complex size distribution dynamics associated with bag breakup, the log-normal distribution, with its inherent long exponential tail, tends to underestimate the size distribution in the case of a bag breakup. This emphasises the crucial significance of selecting a fitting analytical model to ensure accurate predictions of size distribution in scenarios involving bag breakup. The comprehensive analysis conducted in this study provides valuable insights into the efficacy of machine learning architectures for particle/droplet size measurement in the context of digital inline holography and their application in real-world experimental scenarios.

\appendix
\section{Analytical model for droplet size distribution} 
\label{sec:App}

In this segment, we present the volume weights and characteristic sizes related to each mode, \ks{namely,} bag, rim, and node, as described in \cite{jackiw2021aerodynamic} and \cite{jackiw2022prediction}. 

The node breakup volume weight, $w_N$, can be evaluated as follows:
\begin{equation} \label{j:eq4}
w_{N}=\frac{V_{N}}{V_{0}}=\frac{V_{D}}{V_{0}}\frac{V_{N}}{V_{D}},
\end{equation}
where the disk volume $V_D$ is given by \citep{jackiw2021aerodynamic}
\begin{equation} \label{j:eq5a}
\frac{V_{D}}{V_{0}}=\frac{3}{2}\left [ \left ( \frac{2R_{i}}{d_{0}} \right )^{2}\left ( \frac{h_{i}}{d_{0}} \right )-2\left ( 1-\frac{\pi }{4} \right )\left ( \frac{2R_{i}}{d_{0}} \right )\left ( \frac{h_{i}}{d_{0}} \right )^{2} \right ].
\end{equation}

In eq. (\ref{j:eq5a}), $h_i$ represents the thickness of the disk, and $2R_i$ denotes the major diameter of the rim. The expressions for calculating $h_i$ and $2R_i$ are provided as follows: \citep{jackiw2021aerodynamic,jackiw2022prediction}
\begin{equation} \label{j:eq6}
\frac{h_{i}}{d_{0}}=\frac{4}{We_{rim}+10.4},
\end{equation}
and 
\begin{equation} \label{j:eq7}
\frac{2R_{i}}{d_{0}}=1.63-2.88e^{({-0.312We})}.
\end{equation}

In eqs. (\ref{j:eq6}) and (\ref{j:eq7}), $We_{rim}$ denotes the rim Weber number, reflecting the equilibrium between the radial momentum generated at the outer edge of the droplet and the surface tension responsible for droplet stabilisation. The calculation of this parameter is based on the formula $We_{rim}=\rho_{w} \dot{R}^{2}{d_{0}}/\sigma$. The constant rate of radial expansion of a droplet ($\dot{R}$) can be determined as
$\dot{R} = \frac{1.125}{2} \left(\frac{U \sqrt{\rho_a/\rho_w}}{1-32/9 We}\right)$ \citep{jackiw2022prediction}.
The ratio of $V_{N}/V_{D}$ indicates the volume fraction between the nodes and the disk, which is approximately equal to $0.4$ according to \cite{jackiw2022prediction}.

The volume weight of the rim, $w_R$, can be calculated from the following equation \citep{jackiw2021aerodynamic}:
\begin{equation} \label{j:eq5}
w_{R}=\frac{V_{R}}{V_{0}}=\frac{3\pi }{2}\left [ \left ( \frac{2R_{i}}{d_{0}} \right )\left ( \frac{h_{i}}{d_{0}} \right )^{2}-\left ( \frac{h_{i}}{d_{0}} \right )^{3} \right ].
\end{equation}

The volume weight of the bag, $w_B$ is given by:
\begin{equation} \label{j:eq8}
w_B=\frac{V_{B}}{V_{0}}=\frac{V_{D}}{V_{0}}-\frac{V_{N}}{V_{0}}-\frac{V_{R}}{V_{0}}.
\end{equation}

The above-mentioned discussion outlines the procedure for calculating volume weights for each mode. To obtain both individual and overall size distributions, it is crucial to identify the characteristic droplet sizes for each mode. Consequently, the subsequent sections provide an overview of the estimation of these characteristic sizes for each mode.

\subsection*{Node droplet sizes ($d_N$)}
The nodes form along the rim because of the Rayleigh-Taylor (RT) instability, wherein the lighter fluid (air phase) displaces the heavier fluid (liquid phase) \citep{zhao2010morphological}. The droplet size ($d_N$) resulting from node breakup, according to the RT instability theory, is elucidated by \citep{jackiw2022prediction} as follows:
\begin{equation} \label{j:eq9}
\frac{d_{N}}{d_{0}}=\left [ \frac{3}{2}\left ( \frac{h_{i}}{d_{0}} \right )^{2}\frac{\lambda_{RT} }{d_{0}}n \right ]^{1/3},
\end{equation}
where $n=V_{N}/V_{D}$ denotes the volume fraction of nodes relative to the disk. According to \cite{jackiw2022prediction}, the approximated minimum, mean, and maximum values for $n$ are 0.2, 0.4, and 1, respectively. By employing these three $n$ values, the characteristic sizes of the node droplets can be evaluated. Subsequently, the number-based mean and standard deviation for node breakup can be determined based on these characteristic sizes. The maximum susceptible wavelength of the Rayleigh-Taylor instability is provided as, $\lambda_{RT}=2\pi\sqrt{3\sigma/\rho_{w} a}$, such that $a=\frac{3}{4}C_{D}\frac{U^{2}}{d_{0}}\frac{\rho_{a} }{\rho_{w}}\left ({D_{max}/d_{0}} \right )^{2}$ is the acceleration of the deforming droplet. According to \cite{zhao2010morphological}, the drag coefficient ($C_{D}$) of the disk shape droplet is about 1.2, and the extent of droplet deformation is expressed as ${D_{max} / d_0}={2 / (1+\exp{(-0.0019 {We}^{2.7})})}$.
 
\subsection*{Rim droplet sizes ($d_R$)}
The formation of child droplets during rim breakup occurs due to three primary mechanisms: the Rayleigh-Plateau instability, the receding rim instability as elucidated by \cite{jackiw2022prediction}, and the nonlinear instability of liquid ligaments near the pinch-off point.

The size of child droplets ($d_R$) resulting from the Rayleigh-Plateau instability mechanism is as follows:
\begin{equation} \label{j:eq14}
\frac{d_{R}}{d_{0}}=1.89\frac{h_{f}}{d_{0}}.
\end{equation} 
Here, $h_{f}$ is the final rim thickness, which is given by
\begin{equation} \label{j:eq15}
\frac{h_{f}}{d_{0}}=\frac{h_{i}}{d_{0}}\sqrt{\frac{R_{i}}{R_f}},
\end{equation}
where $R_f$ is the bag radius at the time of its burst and it can be evaluated as \citep{kirar2022experimental}
\begin{equation} \label{j:eq16}
R_f=\frac{d_{0}}{2\eta} \left [ 2e^{\tau ^{\prime}\sqrt{p}}+\left ( \frac{\sqrt{p}}{\sqrt{q}}-1 \right )e^{-\tau ^{\prime}\sqrt{q}}-\left ( \frac{\sqrt{p}}{\sqrt{q}}+1 \right )e^{\tau ^{\prime}\sqrt{q}} \right ],
\end{equation}   
where $\eta = f^{2}-120/We$, $p=f^{2}-96/We$ and $q=24/We$. 
In eq. (\ref{j:eq16}), the dimensionless time, denoted as $\tau^{\prime}$, is defined as the ratio of the bursting time ($t_b$) to the characteristic deformation time ($t_d$). These times can obtained as follows \citep{jackiw2022prediction}:
\begin{equation} \label{j:eq17}
t_{b}=\frac{\left [ \left ( \frac{2R_{i}}{d_{0}} \right )-2\left ( \frac{h_{i}}{d_{0}} \right ) \right ]}{\frac{2\dot{R}}{d_{0}}}\left [ -1+\sqrt{1+9.4\frac{8t_{d}}{\sqrt{3We}}\frac{\frac{2\dot{R}}{d_{0}}}{\left [ \left ( \frac{2R_{i}}{d_{0}} \right )-2\left ( \frac{h_{i}}{d_{0}} \right ) \right ]}\sqrt{\frac{V_{B}}{V_{0}}}} \right ],
\end{equation} 
and 
\begin{equation} \label{j:eq18}
t_{d}=\frac{d_0}{U}\sqrt{\frac{\rho_{w} }{\rho _{a}}}.
\end{equation}

The second mechanism contributing to rim breakup is the receding rim instability, as described by \cite{jackiw2022prediction}. This mechanism yields a droplet size ($d_{rr}$) determined as follows:
\begin{equation} \label{j:eq19}
\frac{d_{rr}}{d_{0}}=\left [ \frac{3}{2}\left ( \frac{h_{f}}{d_{0}} \right )^{2}\frac{\lambda _{rr}}{d_{0}} \right ]^{1/3}.
\end{equation}
 Here, $\lambda_{rr}$ denotes the wavelength of the receding rim instability and is determined as $\lambda_{rr} = 4.5b_{rr}$. In this context, $b_{rr}$ represents the thickness of the receding rim and is calculated using the formula $b_{rr} = \sqrt{\sigma / (\rho_{w} a_{rr})}$, where $a_{rr} = U_{rr}^{2}/R_{f}$ characterises the acceleration of the receding rim, as outlined by \cite{wang2018universal}. The receding rim velocity, denoted as $U_{rr}$, is determined experimentally.

The rim breakup occurs as a result of the nonlinear instability of liquid ligaments in proximity to the pinch-off point. To find out the characteristic size associated with this mechanism, it is essential to account for both the Rayleigh-Plateau and receding rim instabilities, as articulated in \citep{keshavarz2020rotary}.
\begin{equation} \label{j:eq20}
d_{sat,R}=\frac{d_R}{\sqrt{2+3Oh_{R}/\sqrt{2}}} ~~{\rm and}
\end{equation}   
\begin{equation} \label{j:eq21}
d_{sat,{rr}}=\frac{d_{rr}}{\sqrt{2+3Oh_{R}/\sqrt{2}}},
\end{equation}
respectively. 
In this context, $Oh_{R}$ is defined as the Ohnesorge number based on the final rim thickness, calculated as $Oh_{R} = \mu / \sqrt{\rho_{w} h_{f}^{3}\sigma}$. The characteristic sizes described in eqs. (\ref{j:eq14}), (\ref{j:eq19}), (\ref{j:eq20}), and (\ref{j:eq21}) are employed to determine the number-based mean and standard deviation for the rim breakup.

\subsection*{Bag droplet sizes ($d_{B}$)}
The droplet size distribution resulting from the rupture of a bag film is influenced by several factors, including the minimum bag thickness, the receding rim thickness ($b_{rr}$), the Rayleigh-Plateau instability, and the nonlinear instability of liquid ligaments. These factors collectively contribute to four characteristic sizes for the satellite droplets, as expressed in \cite{jackiw2022prediction}.
\begin{equation} \label{j:eq21_new}
 d_{B}=h_{min}, 
   \end{equation}  
   \begin{equation} \label{j:eq22}
 d_{rr,B}=b_{rr},
   \end{equation}  
    \begin{equation} \label{j:eq23}
 d_{RP,B}=1.89 b_{rr},
   \end{equation}  
   \begin{equation} \label{j:eq24}
d_{sat,B}=\frac{d_{RP,B}}{\sqrt{2+3Oh_{rr}/\sqrt{2}}}.
   \end{equation} 
\cite{jackiw2022prediction} found that $h_{min} = \pm 2.3$ $\mu \textrm{m}$. Here, $Oh_{rr}$ denotes the Ohnesorge number with respect to the receding rim thickness, indicated as $b_{rr}$. These characteristic dimensions are utilised for determining the number-based mean and standard deviation related to the bag fragmentation mode.
\\

\section{Comparison of droplet size distribution obtained from the traditional method (Hough transform) and the ground truth data}

\begin{figure}[H]
\centering
\includegraphics[width=0.8\textwidth]{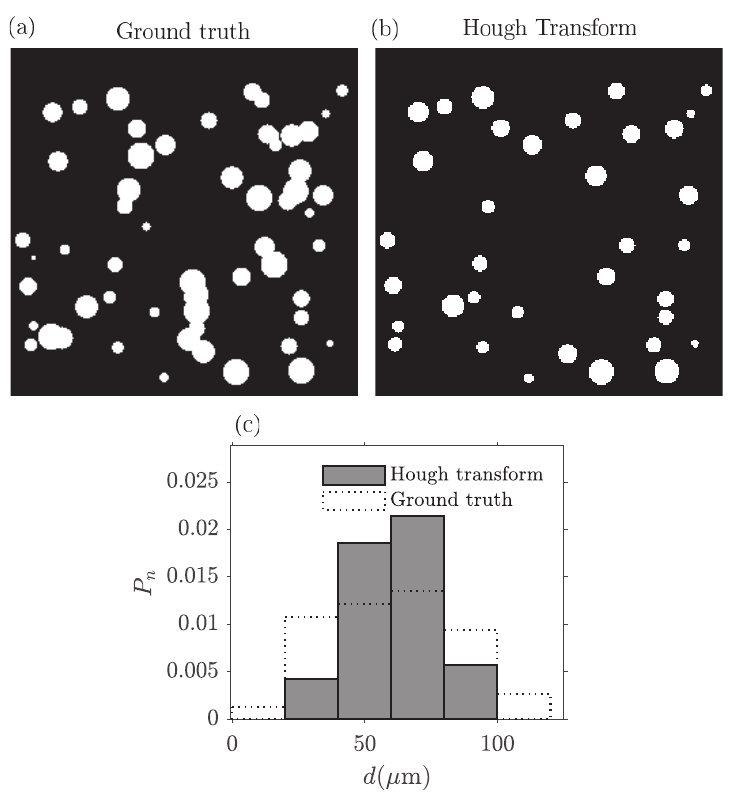}
\caption{\ks{Comparison between the ground truth and the measured distribution from Hough transform for a dataset consisting of 60 droplets randomly distributed in depth. Panels (a) and (b) represent the ground truth image and droplet segmentation image using the Hough transform, respectively. Panel (c) illustrates the size distribution comparison between the hough transform with the ground truth data. Here, $d$ and $P_n$ represent the diameter of the droplets and the number probability density, respectively.}}
\label{fig:drop_comp}
\end{figure}

%\vspace{2mm}
%\noindent{\bf Declaration of Interests:} The authors report no conflict of interest. \\
%\\
\noindent{\bf Acknowledgement:} {K.C.S. thanks the Science \& Engineering Research Board, India, and IIT Hyderabad for their financial support provided through grants CRG/2020/000507 and IITH/CHE/F011/SOCH1.}

%\bibliography{bibl}

\end{document}